\documentclass[aps,prx,twocolumn,superscriptaddress,noeprint,longbibliography]{revtex4-1}

\usepackage[utf8]{inputenc}
\usepackage{amsmath}
\usepackage{amssymb}
\usepackage{wasysym}
\usepackage{graphicx}
\usepackage{hyperref}
\usepackage{xcolor}
\usepackage{physics}
\usepackage{siunitx}

\begin{document}

\title{Floquet prethermalization in a Bose-Hubbard system}

\author{Antonio Rubio-Abadal}
%\email[Electronic address:\,]{antonio.rubio.abadal@mpq.mpg.de}
\thanks{These two authors contributed equally to this work}
\affiliation{Max-Planck-Institut f\"{u}r Quantenoptik, 85748 Garching, Germany}
\affiliation{Munich Center for Quantum Science and Technology (MCQST), 80799 Munich, Germany}
\author{Matteo Ippoliti}
\thanks{These two authors contributed equally to this work}
\affiliation{Department of Physics, Stanford University, Stanford, CA 94305, USA}
\author{Simon Hollerith}
\author{David Wei}
\author{Jun Rui}
\affiliation{Max-Planck-Institut f\"{u}r Quantenoptik, 85748 Garching, Germany}
\affiliation{Munich Center for Quantum Science and Technology (MCQST), 80799 Munich, Germany}
\author{\\S. L. Sondhi}
\affiliation{Department of Physics, Princeton University, Princeton, NJ 08540, USA}
\author{Vedika Khemani}
\affiliation{Department of Physics, Stanford University, Stanford, CA 94305, USA}
\author{Christian Gross}
\affiliation{Max-Planck-Institut f\"{u}r Quantenoptik, 85748 Garching, Germany}
\affiliation{Munich Center for Quantum Science and Technology (MCQST), 80799 Munich, Germany}
\affiliation{Physikalisches Institut, Eberhard Karls Universität Tübingen, 72076 Tübingen, Germany}

\author{Immanuel Bloch}
\affiliation{Max-Planck-Institut f\"{u}r Quantenoptik, 85748 Garching, Germany}
\affiliation{Munich Center for Quantum Science and Technology (MCQST), 80799 Munich, Germany}
\affiliation{Fakultät f\"{u}r Physik, Ludwig-Maximilians-Universität, 80799 Munich, Germany}

\date{\today}

\begin{abstract}
Periodic driving has emerged as a powerful tool in the quest to engineer new and exotic quantum phases. 
While driven many-body systems are generically expected to absorb energy indefinitely and reach an infinite-temperature state, the rate of heating can be exponentially suppressed when the drive frequency is large compared to the local energy scales of the system --- leading to long-lived `prethermal' regimes.
In this work, we experimentally study a bosonic cloud of ultracold atoms in a driven optical lattice and identify such a prethermal regime in the Bose-Hubbard model. 
By measuring the energy absorption of the cloud as the driving frequency is increased, we observe an exponential-in-frequency reduction of the heating rate persisting over more than 2 orders of magnitude. 
The tunability of the lattice potentials allows us to explore one- and two-dimensional systems in a range of different interacting regimes. 
Alongside the exponential decrease, the dependence of the heating rate on the frequency displays features characteristic of the phase diagram of the Bose-Hubbard model, whose understanding is additionally supported by numerical simulations in one dimension. 
Our results show experimental evidence of the phenomenon of Floquet prethermalization, and provide insight into the characterization of heating for driven bosonic systems.
\end{abstract}

\maketitle

\section{Introduction}

The study of out-of-equilibrium dynamics in Floquet systems is an exciting new frontier in quantum physics~\cite{Eckardt2017, Oka2019, Sondhi2020, BukovPolkovnikov2015, MoessnerSondhiReview}.
By driving a quantum system, it is possible to enhance or stabilize interesting equilibrium phases, or even to create new, inherently non-equilibrium phases without a static analogue, such as the discrete time crystal (DTC)~\cite{Khemani2016, else_floquet_2016, CVS, ElseTCReview, SachaReview, Khemani2019} or the anomalous Floquet insulator~\citep{Rudner13, Titum2016, Nathan2019}.
A seemingly ubiquitous obstruction towards realizing such phases in the many-body setting is thermalization: 
by absorbing energy from the drive, a quantum system is generically expected to heat up and eventually approach a featureless ``infinite temperature'' state, which is the maximum entropy state in the absence of any conservation laws~\cite{RigolPeriodicHeating,LazaridesHeating,Ponte15b}. 
The only robust exception to this fate is provided by many-body localization (MBL)~\cite{Anderson58, Basko06, gornyi_interacting_2005, PalHuse, ProsenMBL2008, OganesyanHuse, Luitz15, Imbrie2016, Nandkishore14, EhudMBLRMP, Schreiber2015}, whereby sufficiently strong disorder can prevent this `heat death'~\cite{Lazarides14,Ponte15,Ponte15b,Abanin14, RigolPeriodicHeating,Bordia2017}.
This exception however comes with a number of constraints (e.g. on the presence of disorder, the dimensionality of the system or the range of interactions~\citep{DeRoeck2017, DeRoeckImbrie2017}) that may preclude potentially interesting theoretical scenarios or experimental platforms.

An alternative route towards the realization of non-equilibrium phases is to accept the ultimate thermalizing fate of the system, and focus instead on delaying the inevitable by engineering a long-lived `prethermal' regime. In particular, it has been shown that the timescale for heating can be bounded from below as $t_\text{th} \gtrsim O(e^{\hbar\omega/J_\text{eff}})$ for sufficiently large drive frequency $\omega$, where $J_\text{eff}$ represents a typical local energy scale of the system~\cite{DimaPrethermal_linearresponse, DimaPrethermal_Heff, DimaPrethermal_rigorous, MoriPrethermal1, MoriPrethermal2}. 
Intuitively, such exponentially large heating timescales arise when absorbing one quantum of energy from the drive requires the rearrangement of many local degrees of freedom, which is a high-order process. At times $t\ll t_\text{th}$, the system can in principle exhibit rich dynamics, featuring symmetries, quasi-conserved quantities (including an effective Hamiltonian), etc.~\cite{DimaPrethermal_rigorous, Else2017, Luitz2019}

The existence of prethermal regimes at large driving frequencies was established in a number of analytical and numerical works~\cite{DimaPrethermal_linearresponse, DimaPrethermal_Heff, DimaPrethermal_rigorous, MoriPrethermal1, MoriPrethermal2, Else2017, Luitz2019, Bukov2015, Machado2017, Polkovnikov2018, Luitz2019, RigolDeRoeck2019, Mallayya2019, Haldar2018, Howell2019, Else2019, DeRoeck2019}. On the experimental front, the heating mechanisms of interacting driven systems have lately been the focus of several works based on ultracold atoms in optical lattices, tracking condensate decay, atom losses or doublon production~\cite{Lignier2007, Jotzu2015, Martin2017, Messer2018,Wintersperger2018, Boulier2019}. 
Recently, signatures of prethermalization were observed in the extreme limit of fast, ultra-strong driving~\cite{Weld2019} by measuring the lowest-band depletion.
However, a clear experimental demonstration of frequency-tuned prethermalization, manifested by exponential-in-frequency heating times, has yet to be provided. Such a demonstration requires overcoming several challenges: 
the quantum system, while driven, must preserve coherence for long enough times that the exponential scaling of $t_\text{th}$ with frequency becomes manifest;
at the same time, one must be able to tune $\omega$ across a wide enough range without exciting high-energy degrees of freedom that lie outside the scope of the original isolated system -- e.g. omnipresent higher bands in lattice models;
and, finally, the overall exponential trend must be resolved from other system-specific spectral features that may obscure it in the available frequency range.

\begin{figure*}
\centering
\includegraphics[width=135mm]{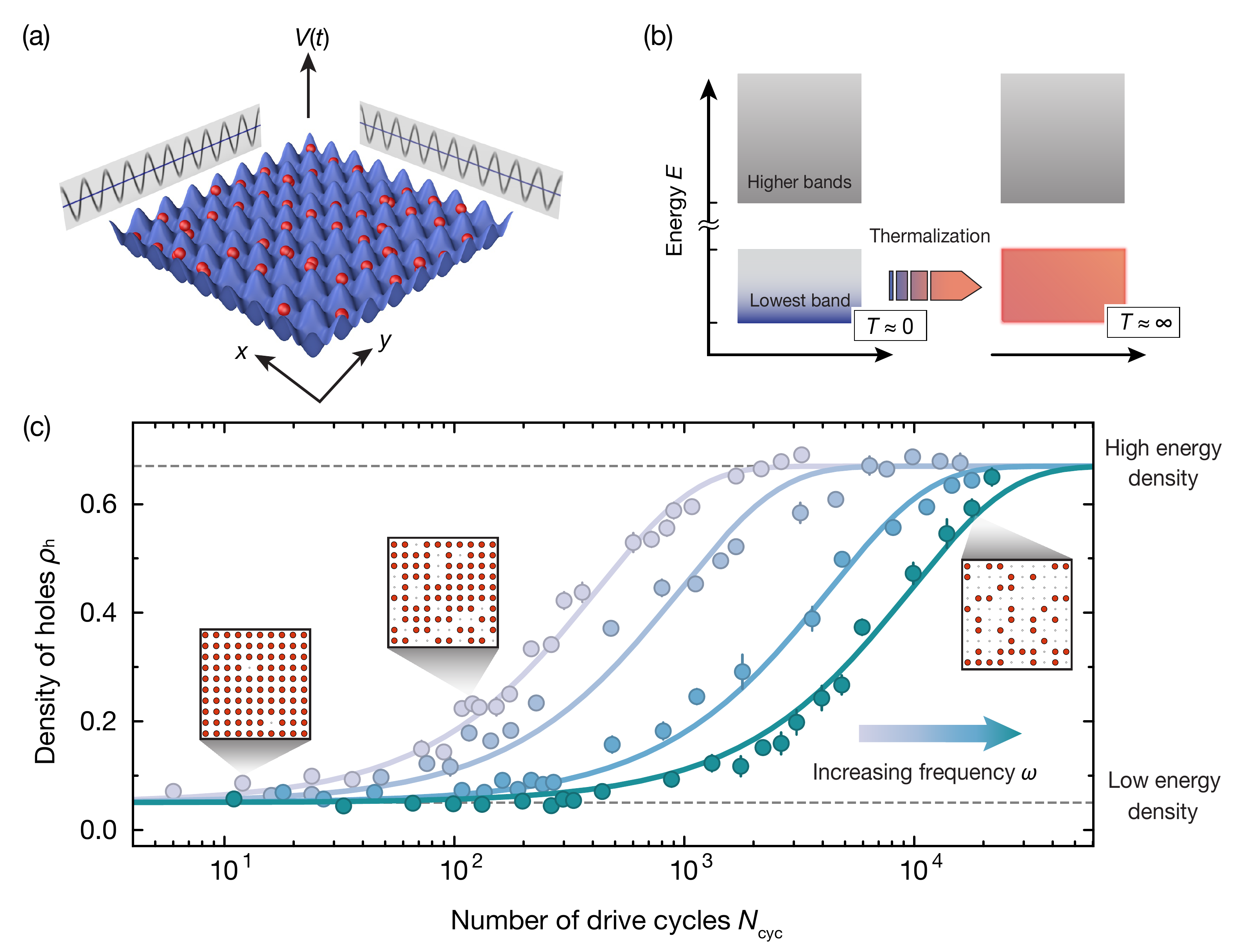}
\caption{ \label{fig:1}
\textbf{Floquet thermalization dynamics in a driven Bose-Hubbard system.} 
(a) Schematic of the two-dimensional system of bosonic atoms in an intensity-modulated optical lattice. The lattice depth $V(t)$ is sinusoidally modulated either in one or two dimensions.
(b) Depiction of the process of Floquet thermalization in our interacting system. 
An initial low-temperature state, involving only the ground state and low-lying excitations, is heated by the drive until it equally populates all many-body states within the lowest band (``infinite temperature'' relative to the lowest band), while higher bands are not populated. 
(c) Dynamics of the density of holes as a function of the driving cycles $N_\text{cyc}$ (measured after ramping the system adiabatically into the atomic limit of a Mott insulator), a proxy for the energy density (see App.~\ref{app:numerics}), in a lin-log plot.   Four different datasets are shown, all driven in the superfluid regime at $J/U=0.16$, for driving frequencies $\omega=19.3 \, J/\hbar$ (gray), $\omega=25.8 \, J/\hbar$ (light blue), $\omega=29.0 \, J/\hbar$ (blue), $\omega=35.5 \, J/\hbar$ (green). The hole density is calculated from a region of interest of $10 \times 10$ lattice sites in the center of the cloud. The solid curves are fits to the exponential form Eq.~\eqref{eq:fitform}. %$\rho_0+(\rho_\infty-\rho_0)(1-e^{-N/N_\text{h}})$. 
The plotted traces display thermalization timescales separated by more than an order of magnitude. The errorbars denote the standard error of the mean (s.e.m.). The square insets show the reconstructed atomic number distribution (red circles), extracted from our raw fluorescence pictures, for three example snapshots.}
\end{figure*}

In this work we show evidence of an exponential-in-frequency thermalization time, the main signature of Floquet prethermalization, in a driven Bose-Hubbard system of ultracold bosonic atoms in a square optical lattice. 
This observation is made possible by two crucial advantages of our experimental setup: 
 (i) the high degree of isolation of our system, which enables us to explore long evolution times~\cite{Choi2016, QuantumBath}; and (ii) a sensitive thermometry technique, enabled by quantum-gas microscopy~\cite{Sherson2010}, which allows us to measure the heating even for weak drives, thereby suppressing transfers to higher bands, and avoiding parametric instabilities. 
 
By tuning the lattice parameters, we explore the response of the atoms in a broad range of couplings spanning the superfluid and Mott-insulating phases. 
The exponentially slow heating is observed most clearly on the superfluid side, where it coexists with weak spectral features possibly associated to Bogoliubov quasiparticle excitations.
On the Mott-insulating side, the heating rate is non-monotonic in frequency, dominated by resonances with higher-occupation excitations (doublons and triplons).
Nonetheless, in both phases, the heating rate drops substantially (by 1 to 2.5 orders of magnitude) even with a modest increase in frequency in the experimentally accessible range. Our understanding of the observed phenomena is aided by numerical simulations in one dimension which, while limited in system size, can explore a broad range of couplings, drive frequencies and timescales.

\section{Experimental setup \label{sec:setup}}

Our experiment begins with the preparation of a two-dimensional cloud of ultracold $^{87}\mathrm{Rb}$ atoms trapped in a single antinode of a vertical optical lattice. 
Next, the cloud is adiabatically loaded into an in-plane square optical lattice at depth $V_0$. We fix the number of atoms such that the density in the central part of the trap is close to one atom per lattice site, typically leading to a total atom number of  $N_\text{at}\simeq 200$. In the prepared state, the atoms only populate the lowest energy band of the lattice potentials, and our system can be well described by a two-dimensional Bose-Hubbard model, with Hamiltonian
\begin{align}\label{eq:dBH}
\hat{H_0} =
-J & \sum_{\langle \textit{\textbf{i}},\textit{\textbf{j}}\, \rangle} \hat{a}_{\textit{\textbf{i}}}^{\dagger} \hat{a}_{\textit{\textbf{j}}}^{\mathstrut}
+ \frac{U}{2} \sum_{\textit{\textbf{i}}} \hat{n}_{\textit{\textbf{i}}}(\hat{n}_{\textit{\textbf{i}}}-1) 
+ \sum_{\textit{\textbf{i}}}  \epsilon_\textit{\textbf{i}}\, \hat{n}_{\textit{\textbf{i}}}, 
\end{align}
where $\hat{a}_{\textit{\textbf{i}}}^{\mathstrut} \,$, 
$\hat{a}_{\textit{\textbf{i}}}^{\dagger}$ and 
$\hat{n}_{\textit{\textbf{i}}}^{\mathstrut} $ respectively denote the annihilation, creation and number operators at a site $\textit{\textbf{i}}$ of the square lattice [$\textit{\textbf{i}}=(i_{x},i_{y})$], 
$J$ is the tunneling amplitude between nearest-neighbour sites $\langle \textit{\textbf{i}},\textit{\textbf{j}} \,\rangle$, $U$ the on-site interaction energy, and $\epsilon_{\textit{\textbf{i}}} $ the harmonic trapping potential (see Appendix \ref{app:exp}).

At this stage, the atoms are at very low temperature, close to the ground state of Eq.~\eqref{eq:dBH}. 
We then start driving the system by periodically modulating the depth of the in-plane lattices as $V(t)=V_0(1+A \cos( \omega t))$ (see Fig.~\ref{fig:1}(a)), where $A$ is the relative modulation amplitude, leading to a time-dependent modulation of all the Hamiltonian parameters. However, the tunneling strength, due to its exponential dependence on the lattice depth, dominates the modulation so that
\begin{align} 
\hat{H}(t)\approx \hat{H}_0+g\cos(\omega t) \hat{O}_\text{drv} \;,
\label{eq:timedepH}
\end{align}
with $g = \delta J/J$ and $\hat{O}_\text{drv} = J \sum_{\langle \textit{\textbf{i}},\textit{\textbf{j}}\, \rangle} \hat{a}_{\textit{\textbf{i}}}^{\dagger} \hat{a}_{\textit{\textbf{j}}}^{\mathstrut}$~\cite{Stoferle2004} . 
To ensure that during the driving no atoms are excited into higher bands, we keep the driving frequencies well below the bandgap~\cite{Sun2018} and use a low modulation amplitude $A\ll 1$ to avoid multi-photon interband transitions~\cite{Weinberg2015} (see Appendix~\ref{app:exp}). 

After driving the system for an integer number of periods, $N_\text{cyc}=\omega\, t_\mathrm{drv} /2 \pi $, we slowly ramp up the lattice depth until the system becomes an atomic-limit Mott insulator. 
At this stage, all tunneling dynamics is frozen and, if no heating took place during the drive, this results in a Mott insulator near unit filling.
Finally, we measure the parity-projected atomic occupation of the lattice sites through fluorescence imaging~\cite{Sherson2010}. 
Because of parity projection, the growth of the hole density is directly linked to excitation processes that
increase the variance of the single-site occupation. 
Thus, the density of holes is a proxy for the energy density of the cloud and thus for the heating dynamics (see App.~\ref{app:numerics}). 

\section{thermalization dynamics \label{sec:expdyn}}

In Fig.~\ref{fig:1}(c), we plot the evolution of the density of holes $\rho_\text{h}$ under driving for four different frequencies. The depth of the in-plane lattices is $V_0 =8\,E_\mathrm{r}$, where $E_\text{r}=h^2/8ma_\text{lat}^2$ is the recoil energy, with $m$ being the atomic mass and $a_\text{lat}$ the lattice spacing. The relative modulation amplitude of the periodic drive is $A=0.05$. 
While each drive frequency in Fig.~\ref{fig:1}(c) gives rise to a qualitatively similar time dependence of $\rho_\text{h}$, the thermalization times themselves are vastly different -- spanning 1.5 orders of magnitude within less than a factor of 2 in $\omega$.  
Such a striking dependence on the drive frequency is an indication of the exponential slow-down of heating characteristic of Floquet prethermalization. 

Theoretically, the phenomenon of prethermalization refers to the quasi-conservation of energy stemming from the existence of a (quasi-) \emph{local} time-independent Hamiltonian that captures the dynamics of the system out to an exponentially long time. 
While one can always formally define a (non-unique) `Floquet Hamiltonian' from the unitary time evolution operator via $\hat{U}(T) = \mathcal{T} e^{-i\int_0^T {\rm d}t\, \hat{H}(t)/\hbar} \equiv e^{-i \hat{H}_F T/\hbar}$, the operator $\hat{H}_F$ is generally highly non-local in a many-body system (and hence unphysical). 
Nevertheless, when the frequency $\omega$ is large compared to the local energy scales of the problem (here denoted collectively by $J_\text{eff}$), one can perform a high-frequency asymptotic expansion for $\hat{H}_F$ in powers of $1/\omega$, $\hat{H}_F = \sum_n (1/\omega)^n \, \hat{\mathcal{H}}_F^{(n)}$;
the leading-order term $\hat{\mathcal H}_F^{(0)}$ equals the time-averaged Hamiltonian, while higher-order terms are progressively longer-ranged and contribute significantly to the dynamics only at correspondingly later times. 
While ultimately divergent, this expansion looks convergent out to some optimal order $n_\text{opt} = O(\omega/J_{\rm eff})$.
Truncating the expansion at this order yields an exponentially accurate approximation to the Floquet time evolution $\hat{U}(T)$, which sets the rate of heating to be exponentially small~\cite{DimaPrethermal_linearresponse, DimaPrethermal_rigorous, MoriPrethermal1, MoriPrethermal2}, giving
\begin{equation}
E(N_\text{cyc})/N_{\rm at} \lesssim N_\text{cyc}\, J_\text{eff}\, e^{-\hbar \omega/J_{\rm eff}} +O\left(1/\omega \right) \;,
\label{eq:Etbound}
\end{equation}
where $E(N_\text{cyc}) = \langle \hat{\mathcal H}_F^{(0)}\rangle_{N_\text{cyc}}$ is the energy absorbed by the system.

Thus, in our experiment one expects the energy density to at first increase linearly in time (as is also expected, e.g., from linear response theory), before eventually saturating to its infinite-temperature value;
other local observables such as $\rho_\text{h}$ are expected to follow the same behavior.
While more intricate behavior could in principle arise at intermediate times, the simplest ansatz for the stroboscopic time dependence of $\rho_\text{h}$ takes the form 
\begin{equation}
\rho_{\text{h}} (N_\text{cyc}) \simeq \rho_0+(\rho_\infty-\rho_0)\left[ 1-\exp(-N_\text{cyc}/N_\text{cyc}^\text{th})\right] \;,
\label{eq:fitform}
\end{equation}
where $\rho_0$ is the baseline value measured in the absence of the drive, $\rho_\infty$ is the infinite-temperature value, and $N_\text{cyc}^\text{th}$ is the heating timescale, predicted to obey $N_\text{cyc}^\text{th} > e^{O(\omega)}$ for sufficiently high $\omega$.

Fits to the data of the form~\eqref{eq:fitform} (solid lines in Fig.~\ref{fig:1}(c)) in fact reveal a good agreement, and yield thermalization timescales $N_\text{cyc}^\text{th}$ between $4\times 10^2$ at the lowest frequency and $10^4$ at the highest. Note that the long times measured in our data, over $3000\, \hbar/J$, are crucial to the detection of this effect and reflect the high level of isolation of our system, in which any intrinsic decoherence processes are highly suppressed (see Appendix~\ref{app:exp}). We remark that what may look like a plateau in $\rho_\text{h}$ at early times is in fact a very slow linear growth as follows from Eq.~\ref{eq:Etbound} (and as is true more generally of prethermal ``plateaux'').
A nonequilibrium initial state, e.g. one with a spatial density imbalance, would first thermalize relative to the leading-order prethermal Hamiltonian (in our case $\hat{\mathcal{H}}_F^{(0)}\approx \hat{H}_0$, see Eq.~\eqref{eq:timedepH}) on a timescale independent of $\omega$, and then heat to infinite temperature exponentially slowly. 
However, in our experiment we do not observe any fast transient dynamics before the onset of this slow heating because the initial state we prepare is \emph{already} in thermal equilibrium relative to $\hat{\mathcal{H}}_F^{(0)}$.

While the dynamics of $\rho_\text{h}$ illustrate qualitatively the phenomenon of Floquet prethermalization, a more precise characterization of the heating-rate dependence can be obtained from a fit of the atomic density profile. This established thermometry method, based on the fit of a grand-canonical model~\cite{Sherson2010, Endres2012}, provides us with a measure for the temperature of the cloud, from which we can characterize the heating induced by the drive. Due to the high sensitivity of this technique, we can better explore the linear response regime, where the drive amplitude $A$ is small such that interband processes are strongly suppressed (see Appendix \ref{app:lr_exp}). 
This weak-drive probing is in contrast to recent measurements of the response of Bose-Einstein condensates in one and two-dimensional optical lattices~\cite{Boulier2019, Wintersperger2018}, which focused instead on the emergence of parametric instabilities under strong drives.

\section{Heating in the Bose-Hubbard model: Numerics \label{sec:numerics}}

Before moving on to the results of the experiment outlined above, it is useful to gain some intuition on the nature of Floquet heating in the Bose-Hubbard phase diagram with the help of numerical simulations.
A variety of methods have been applied to the Bose-Hubbard model out of equilibrium~\cite{Kennett2013, Huber2007, Strand2015, Eckardt2017, Cirac2019, Mallayya2019}, with particular interest on parametric instabilities of the superfluid condensate in recent years~\cite{Bukov2015, Lellouch2017}.
To study the approach to infinite temperature under weak driving, we use numerical exact diagonalization and the Krylov subspace method for time evolution~\cite{Manmana2005} as detailed in the following.

While the general theory of prethermalization applies to arbitrarily strong drives, our experiment considers a weak modulation $g\ll 1$. 
In this regime, the energy absorbed per Floquet cycle by the system is well captured within linear response theory as the dissipative part of a response function,
\begin{equation}
%N_h^{-1} \sim 
\Phi(\omega) = \sum_{n\neq 0} | \langle n | \hat{O}_{\rm drv} | 0 \rangle |^2 \, \delta(E_n-\hbar \omega) \;, \label{eq:LR}
\end{equation}
where $\{ E_n, |n\rangle\}$ label the eigenvalues and eigenvectors of the time-averaged Hamiltonian $\hat{H}_0$ in Eq.~\eqref{eq:dBH} ($|0\rangle$ being the ground state, with $E_0 = 0$),
and $\hat{O}_\text{drv}$ has been introduced in Eq.~\eqref{eq:timedepH}. 
We note that $\Phi(\omega)$ is also the quantity rigorously bounded by an exponential in Ref.~\citep{DimaPrethermal_linearresponse} 
(the result there is proven for systems with bounded energy density, e.g. fermions on a lattice, but we argue in Appendix~\ref{app:ptb} that a slightly relaxed version applies to the present case of bosons near unit filling as well). $\Phi(\omega)$ has units of energy and, for weak drives ($g\ll 1$), is proportional to the energy absorbed per Floquet cycle, $ dE/dN_{\rm cyc} \sim g^2 \Phi(\omega)$. Strictly speaking it quantifies the rate at which the ground state is depleted, though in the following we will refer to $\Phi(\omega)$ as a `heating rate' for simplicity.

\begin{figure}
\centering
\includegraphics[width=\columnwidth]{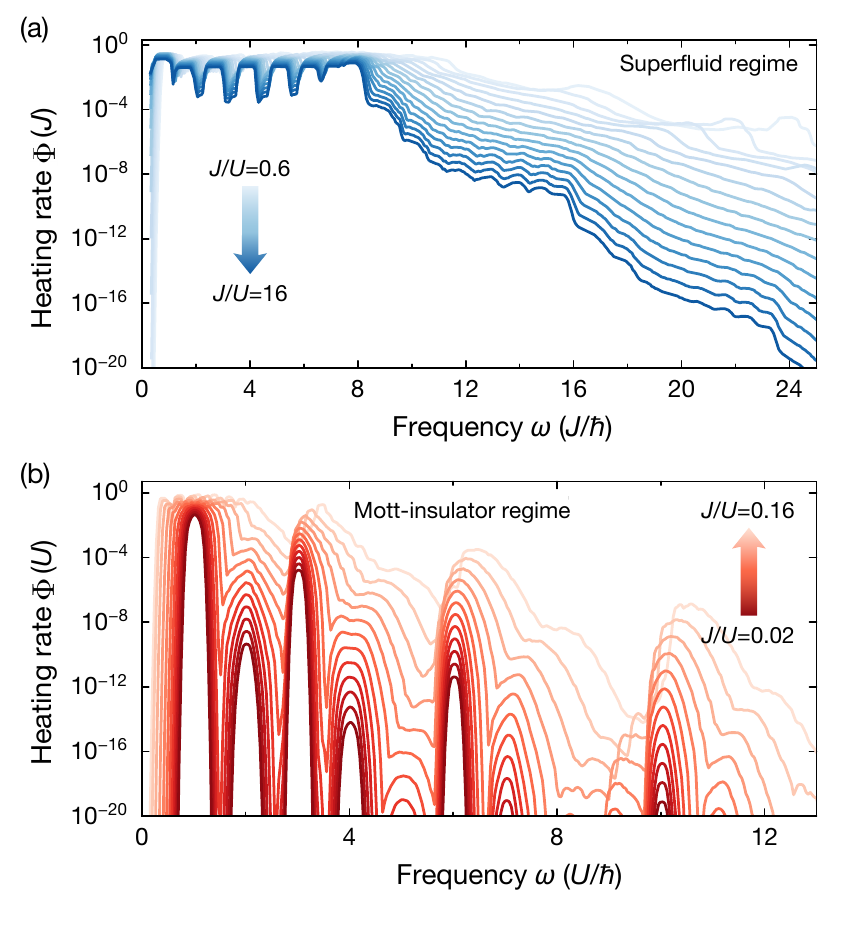}
\caption{\label{fig:numerics} \textbf{Numerical simulation of heating in the 1D Bose-Hubbard model.} 
The linear response heating rate $\Phi(\omega)$, computed by numerical exact diagonalization of a chain of $L=9$ sites at unit filling, as a function of drive frequency $\omega$ in (a) the superfluid and (b) the Mott-insulating phase. 
In (a), $J/U$ varies from 0.6 (lightest blue) to $16$ (darkest blue), while in (b) it varies from $0.02$ (darkest red) to 0.16 (lightest red).
The critical point is at approximately $J/U = 0.26$.}
\end{figure}

We compute $\Phi(\omega)$ in Eq.~\eqref{eq:LR} via numerical exact diagonalization of a one-dimensional chain of $L=9$ sites at unit filling. 
Given the small system size, we replace the harmonic trap potential in $\hat{H}_0$ with hard-wall (open) boundary conditions
(for additional details on the numerics see Appendix~\ref{app:numerics}).
This approach, while strongly limited in system size, provides complete flexibility in the choice of couplings $J/U$ and frequency $\omega$, while also allowing us to probe extremely long timescales (within linear response).
The results, shown in Fig.~\ref{fig:numerics}, outline a clear picture of the nature of heating in the two phases.
To best highlight each phase's spectral features, we show the heating and the frequency in units of $J$ in the superfluid phase (see Fig.~\ref{fig:numerics}(a)), and $U$ in the Mott insulating phase (see Fig.~\ref{fig:numerics}(b)).

Deep in the superfluid phase, the system manages to heat efficiently up to frequency $\omega = 8J/\hbar$, then the rates sharply drop in an exponential fashion, with additional kink-like features visible at $\omega = 16J/\hbar$ and higher multiple frequencies. 
This suggests that heating takes place via the excitation of quasiparticles from the condensate. 
As the drive carries no net momentum, the quasiparticles must come in pairs with opposite momenta $\pm q$. 
For $\omega > \Omega_{\text{2qp}}$ (twice the quasiparticle bandwidth), to absorb a single quantum of energy $\hbar \omega$ from the drive, the system must concurrently scatter multiple pairs from the condensate into excited states, with each additional scattering event suppressing the amplitude by factors of $U/J \ll 1$.
This gives the expected exponential scaling and explains the threshold features visible in Fig.~\ref{fig:numerics}(a) at frequencies commensurate with $8J/\hbar$ (twice the single-particle bandwidth). 
Increasing the interaction strength $U$ gradually washes out the above features, 
while pushing the value of $\Omega_{\text{2qp}}$ upwards approximately in agreement with the Bogoliubov prediction $\Omega_{\text{2qp}} \simeq 8J/\hbar \sqrt{1+U/2J}$ -- though notice the latter is a mean-field prediction and as such it is expected to work better in higher dimension.

As $U$ increases further, eventually the superfluid spectral features give way to sharp peaks associated to higher on-site occupancy (doublons, triplons, etc.), a characteristic of the Mott-insulating phase.
In fact, Fig.~\ref{fig:numerics}(b) displays a hierarchy of peaks at integer values of $\hbar \, \omega/U$. The location and height of each peak can be understood by perturbing away from an atomic limit ($J=0$) Mott state (see Appendix~\ref{app:ptb}). 
While the envelope of the peaks does obey an exponential bound (in one dimension -- see Appendix~\ref{app:ptb} for a discussion of higher dimension), the strongly non-monotonic structure of $\Phi(\omega)$ means that measurements with a limited dynamic range in $\omega$ and $N_\text{cyc}^\text{th}$ may not be able to identify the overall trend. 

While these results have been obtained for the ground state of the system, the exponential suppression also holds at finite temperatures. In Appendix~\ref{app:numerics} we show that considering an ensemble at intermediate and high temperatures does not lead to significant changes in the results we present in this section.
In addition, we also complement the analysis in this section by studying the approach of the system to infinite temperature via exact time evolution.
This confirms that the above picture remains valid even beyond the linear response regime, and shows that the zero-temperature spectral function $\Phi(\omega)$ captures the heating timescales remarkably well.

%The heating dynamics generally takes this form only at late times, close to infinite temperature, where $\kappa$ becomes a temperature-independent constant~\cite{Mallayya2019}. 

\section{Experimental results: Heating in two dimensions}

\begin{figure*}
\centering
\includegraphics[width=168mm]{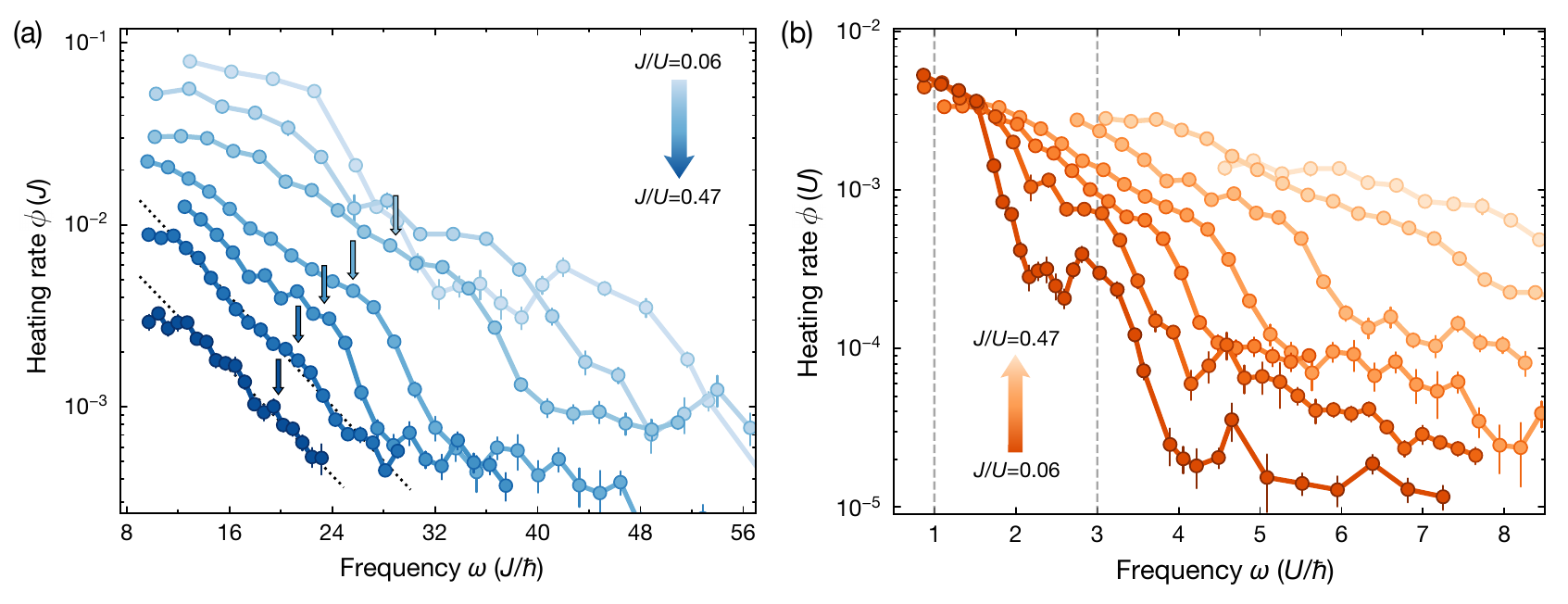}
\caption{ \label{fig:2d}
\textbf{Heating rate per Floquet cycle in 2D vs drive frequency.} The heating rates and driving frequencies are expressed in units of (a) the tunneling strength $J$ and (b) the on-site interaction $U$. The range of lattice depths $V_0$ varies from $5 -11\,E_\mathrm{r}$ in steps of $1\,E_\mathrm{r}$ (correspondingly $J/U$ varies from $0.47-0.06$). In (a), the datasets go from shallower (dark blue) to deeper (light blue) lattices. Five small arrows indicate the position of $\Omega_{\text{2qp,2D}}$ for the first four datasets (see main text). The two dotted lines are exponential fits of the first two datasets. In (b), the datasets go from deeper (dark orange) to shallower (light orange) lattices. The dashed vertical gray lines indicate the doublon and triplon resonances at $\omega=U/ \hbar$ and $3\,U/ \hbar$ respectively. The errorbars denote the s.e.m.}
\end{figure*}

We now turn to experimentally characterizing the slow Floquet thermalization, by extracting the temperatures from the density profile of the cloud, in our two-dimensional system. We measure the heating rates at different lattice depths within $V_0=5-11E_\mathrm{r}$ over a range of driving frequencies with a fixed relative modulation  $A=0.05$. These heating rates are extracted from a linear fit of the temperature (see Appendix~\ref{app:lr_exp}), and we express them as the energy absorbed per Floquet cycle $\phi(\omega)$, which is related to $\Phi(\omega)$ through $\phi(\omega)= (\pi g)^2 \Phi(\omega)$.
The results (see Fig.~\ref{fig:2d}(a)) reveal a clear suppression of the heating rate as the frequency is increased, extending over more than two decades in the measured range. 
This abrupt arrest of Floquet thermalization manifests the presence of a Floquet prethermal regime.

For values of $J/U$ well above the phase transition at  $J/U\simeq 0.06$~\cite{Capogrosso2008}, i.e. in the superfluid phase,  all datasets show qualitatively the same behavior -- a monotonic and approximately exponential decrease of the heating rate as the frequency is increased. This trend is further indicated, for the two datasets with highest $J/U$, by the fit of an exponential function $\phi(\omega)=C \, e^{-\hbar\omega/J_\text{eff}}$ (see dotted lines in Fig.~\ref{fig:2d}(a)). The fit allows us to extract the effective local energy scales, $J_\text{eff,1}=5.76(16)\,J$ and $J_\text{eff,2}=5.9(2)\,J$, which are consistently on the same order of magnitude as the Hamiltonian parameters $J$ and $U$. As we move away from weak interactions, a visible kink-like feature appears on top of the general exponential trend.  
Following the line of argument from the previous section, we expect the dominant heating process in the superfluid to be the generation of quasiparticle pairs with opposite momenta. 
Because of this, the heating is expected to be further reduced for drive frequencies above twice the Bogoliubov quasiparticle bandwidth, which in the 2D case is $\Omega_\text{2qp,2D}=2 \times 8 J/\hbar \sqrt{1+U/4J}$~\cite{Bukov2015}. 
In Fig.~\ref{fig:2d}(a) we indicate with five small arrows the frequency $\Omega_\text{2qp,2D}$ for the five traces with weakest interactions, which roughly agree with the positions of the kinks observed in the data. 

Aside from this feature, we note that below twice the non-interacting bandwidth (here $16J/\hbar$) the heating rate is not flat, in contrast to the results for the 1D numerics.
This is to be expected based on the different density of states of a tight-binding model in a square lattice in 1D and 2D: in the latter case, the density of states has a maximum in the middle of the band, which means the excitation of quasiparticles is most efficient, and the heating fastest, near twice the middle of the bandwidth (here $\omega = 8J/\hbar$)~\cite{Huber2008}.
Finally, we also note that the dynamic range of the driving frequencies is smaller for higher $J/U$, since higher tunneling strengths $J$ require higher frequencies (on an absolute scale) to see prethermalization, making the limitation posed by interband heating more severe. 

Moving to the strongly interaction regime, different features emerge. The dataset with the smallest $J/U$, in fact the only one in the Mott-insulating phase, shows a distinct non-monotonic behavior in the observed frequency range. To identify the relevant spectral features associated with it, in Fig.~\ref{fig:2d}(b) we show the same data as in (a) but expressed in units of the interaction strength $U$. Indeed the trace at $J/U=0.06$ (dataset in dark orange) shows a peaked structure at $\omega=U/\hbar$ and  $3\, U/\hbar$, as expected for the doublon and triplon resonances respectively. As the interaction strength is reduced, these resonant features fade into a continuum associated with the superfluid bandwidth, similar to what one observes for the numerics in Fig.~\ref{fig:numerics}(b).

Finally, we also note that in the regime of very high frequencies, which features the smallest heating rates, we reach the sensitivity limit of this experiment. This is caused by the very long measurements times and the then significant contribution of the background heating present in our system. The noise floor is expected to be dependent on $J/U$ due to the change in the excitation spectrum of the system.

\section{Experimental results: Heating in one dimension}

\begin{figure}
\centering
\includegraphics[width=80mm]{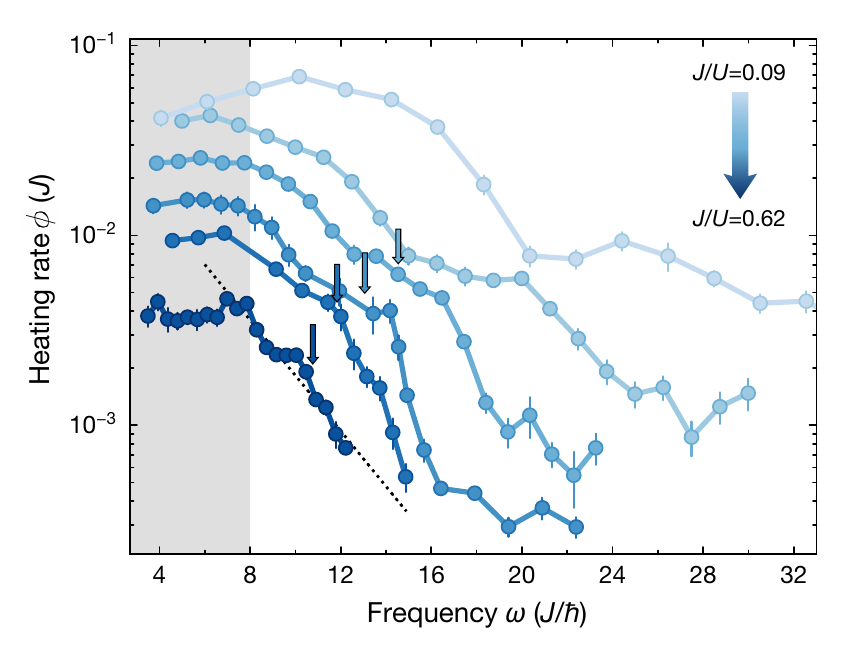}
\caption{ \label{fig:1d}
\textbf{Heating rate per Floquet cycle in 1D vs drive frequency.} The heating rates and driving frequencies are expressed in units of the tunneling strength $J$. The datasets vary from shallower (dark blue) to deeper (light blue) lattices with lattice depths $V_{0,x}= 3, 4, 5, 6, 7, 9 E_\mathrm{r}$. The values of $J/U$ vary correspondingly from $0.62-0.09$. Four small arrows indicate the position of $\Omega_{\text{2qp,1D}}$ for the first four datasets (see main text). The dotted line indicates an exponential fit of the first dataset.The grey shaded are indicates the region within twice the single-particle bandwidth, $\omega=8J/ \hbar$. The errorbars denote the s.e.m.}
\end{figure}

Our experimental setup allows us also to produce 1D systems. 
We achieve this by ramping the lattice along the $y$ axis to a depth of $V_{0,y} = 20 E_\mathrm{r}$ before the start of the drive.  The typical atom number in each one of these 1D systems is of $N_\text{at} \simeq 15$.
The measured heating rates in this 1D geometry are shown in Fig.~\ref{fig:1d}, displayed in a similar fashion as in the 2D case, for lattice depths at $V_{0,x}=3-9E_\mathrm{r}$ and with a relative lattice modulation of $A_x=0.1$ (while $A_y=0$).
Here too we observe an exponential suppression of the heating rate as a function of the drive frequency $\omega$.
However, for $\omega<8J/\hbar$ the heating rate appears almost constant, in agreement with the numerics in Fig.~\ref{fig:numerics}(a), and only beyond this flat part we see a sharp decrease. 
While this behavior can be solely explained in terms of twice the \emph{non-interacting} bandwidth of the system, $2 \times 4J/\hbar$, we also observe a second kink at slightly higher frequency, reminiscent of the 2D case, which shifts to higher frequencies as interactions increase. 
In Fig.~\ref{fig:1d}  we also use four arrows to indicate the position of twice the Bogoliubov bandwidth, $\Omega_\text{2qp,1D}=2 \times 4 J/\hbar \sqrt{1+U/2J}$, for the first four datasets, showing a reasonable agreement that eventually becomes discrepant for higher interactions. We also include an exponential fit of the first dataset, with $J/U=0.62$, taken only above $\omega<8J/\hbar$ and from which we extract an effective local energy scale of $J_\text{eff}=3.0(3)\,J$. Note that in this case the presence of interactions leads already to a deviation from the simple exponential trend, even for the weakest interaction explored here.

For stronger interactions, the heating rate becomes nonmonotonic, analogously to the 2D results and the 1D numerics, though in this case the associated features are less sharp. 
This could be explained by the inhomogeneity present in the 1D system, due to much stronger confinement from the transverse lattice.
We also note that, while the numerical simulations capture most features of the 1D experimental data,  quantitative discrepancies between the timescales are to be expected for various reasons, chiefly the different boundary conditions (hard wall vs harmonic confinement), but also the drive amplitudes (infinitesimal vs finite) and protocols (the experiment naturally includes a weak modulation of $U$ which is not considered in the simulations).

\section{Conclusion}

We have studied the nature of heating in a system of driven ultracold bosonic atoms in an optical lattice, and found strong evidence that the thermalization time diverges exponentially in the drive frequency -- a central prediction in the theory of Floquet prethermalization. When increasing the frequency only by a factor of 2 or 3, we were able to suppress the heating rate by as much as two orders of magnitude. This possibility of driving isolated quantum systems while avoiding heating for exponentially long times is a key step towards the engineering of new forms of matter existing only out of equilibrium.

Furthermore, our results add physical insight to the intuitive picture of Floquet prethermalization -- that a quantum system driven at high enough frequency heats slowly because the absorption of a quantum of energy $\hbar \omega$ from the drive requires many coordinated rearrangements of its local degrees of freedom. 
Our experimental results not only confirm this picture, but also shed light on the understanding of what those rearrangements look like in a real system. 
Interestingly, they point to different scenarios for the superfluid and Mott-insulating sides of the Bose-Hubbard phase diagram.

In the future, it will be interesting to explore Floquet heating in our system in the limit of hard-core bosons, where a different prethermalization mechanism, based on weak integrability breaking, may be realized~\cite{Mallayya2019}.
Another direction is the addition of disorder, where it would be interesting to microscopically characterize the failure of the MBL phase to heat to infinite temperature. 
Other possible directions include implementing more exotic drives, such as quasiperiodic ones~\cite{Else2019, DeRoeck2019};
using strong drives to probe the heating rates with our technique beyond the linear response regime;
and exploring the dependence on the initial temperature. Most interestingly, our work paves the way for future realizations of novel prethermal Floquet phases of matter with no equilibrium analogs. 

\textit{Note added.} During the preparation of this manuscript, we became aware of related work providing evidence for the observation of exponential-in-frequency thermalization times in dipolar spin chains~\citep{Cappellaro2019}. 

\textit{Correspondence address:} \href{mailto:antonio.rubio@mpq.mpg.de}{antonio.rubio@mpq.mpg.de}.

\begin{acknowledgments}
The authors would like to acknowledge useful discussions with Emanuele G. Dalla Torre, Sarah Hirthe, Michael Knap, Roderich Moessner, Marcos Rigol, Nepomuk Ritz, Alexander Schuckert, Dan M. Stamper-Kurn and David Weld. We also thank Simon Evered and Kritsana Srakaew for careful reading of the manuscript. 
This work was supported with funding from the Defense Advanced Research Projects Agency (DARPA) via the DRINQS
program. The views, opinions and/or findings expressed are those of the authors and should not be interpreted as
representing the official views or policies of the Department of Defense or the U.S. Government.
We also acknowledge funding by the Max Planck Society (MPG), the European Union (PASQuanS Grant No. 817482) and the Deutsche Forschungsgemeinschaft (DFG, German Research Foundation) under Germany’s Excellence Strategy – EXC-2111 – 390814868. M.I. was funded in part by the Gordon and Betty Moore Foundation's EPiQS Initiative through Grant GBMF4302 and GBMF8686.
 J.R. acknowledges funding from the Max Planck Harvard Research Center for Quantum Optics.
\end{acknowledgments}

\appendix

\section{Experimental details and parameters \label{app:exp}}
\subsection{Description of the setup}

After the preparation of a two-dimensional degenerate gas of Rubidium-87 atoms, we load the atoms into a square optical lattice, generated by two retroreflected laser beams in the atomic plane, with lattice spacing $a_\text{lat}=\SI{532}{nm}$. We ramp up the in-plane optical lattices with an s-shaped ramp of $\SI{75}{ms}$ up to a lattice depth $V_0$ expressed in units of the energy recoil $E_\text{r}=h^2/8ma_\text{lat}^2$, where $m$ is the atomic mass. Ramps of the same duration and form are used to ramp to the transition point and to the atomic limit after the driving dynamics, where the energy content of the gas is measured.

In addition to the lattice potential, the atoms experience an overall harmonic trapping potential, given by $\epsilon_\textit{\textbf{i}}=m a^2_{\text{lat}}(\omega^2_x i_x^2+\omega^2_y i_y^2)/2$, where $\omega_x$ and $\omega_y$ are the harmonic trap frequencies in the plane, which are slightly different for each lattice depth. In the 2D case, they are in the range  $2\pi \times \SI{ 45}{Hz}< \omega_x$,$\omega_y < 2\pi\times \SI{55}{Hz}$, while in the 1D case $ \omega_x \simeq  2\pi\times \SI{ 70}{Hz}$ is roughly constant.

\subsection{Bose-Hubbard parameters}
The values of $J$ and $U$ used in the main text are obtained from a numeric calculation of the band structure, and are based on the calibrated lattice depths $V_0$, which are estimated to have an uncertainty of roughly $2\%$. 
We show here all the calculated parameters for the relevant depths corresponding to Fig.~\ref{fig:2d} and Fig.~\ref{fig:1d}.
In the 2D case, both in-plane lattices had the same lattice depth and the lattice modulation was $A=0.05$ (see Tab.~\ref{tab:2D}). 
In the 1D case, the lattice along the y axis was fixed to $20 E_\text{r}$ and the lattice along the $x$ axis was tuned to $V_{0, x}$ and driven with amplitude $A_x=0.1$ (see Tab.~\ref{tab:1D}). 
We also plot the modulation of the tunneling strength $\delta J$, obtained as $\delta J=(J_{V_0-A}-J_{V_0+A})/2$.

\begin{table}[h]
\begin{tabular}{|c|c|c|c|c|c|}
\hline
$V_0$($E_\text{r}$) & $J/h$(Hz) &  $U/h$(Hz) & $J/U$ & $\delta J/h$(Hz) & $\delta J /J$  \\ \hline
5       & 134.0      & 288 & 0.47   & 8.8  & 0.067    \\ \hline
6       & 103.2      & 327 & 0.32   & 8.0  & 0.077    \\ \hline
7       &   80.0    & 363 & 0.22   & 7.0  & 0.088    \\ \hline
8       & 62.5      & 396 & 0.16   & 6.1  & 0.098    \\ \hline
9       & 49.2      & 427 & 0.11   & 5.3  & 0.11    \\ \hline
10      & 38.9     & 457 & 0.085   & 4.5  & 0.12    \\ \hline
11      & 31.0     & 485 & 0.064   & 3.8  & 0.12    \\ \hline
\end{tabular}
\caption{ \label{tab:2D}
\textbf{Table of Bose-Hubbard parameters for the 2D experiment.}}
\end{table}

\begin{table}[h]
\begin{tabular}{|c|c|c|c|c|c|}
\hline
$V_{0, x}$($E_\text{r}$) & $J_x/h$(Hz) &  $U/h$(Hz) & $J_x/U$ & $\delta J_x/h$(Hz) & $\delta J_x /J_x$  \\ \hline
3       & 229.1      & 288 & 0.62   & 18.7  & 0.08    \\ \hline
4       & 174.9      & 412 & 0.42   & 18.8  & 0.11     \\ \hline
5       & 134.0      & 446 & 0.30   & 17.7  & 0.13    \\ \hline
6       & 103.2      & 475 & 0.22   & 16.0  & 0.16    \\ \hline
7       &   80.0    & 500 & 0.16   & 14.1  & 0.18    \\ \hline
9       & 49.2      & 543 & 0.09   & 10.6  & 0.22    \\ \hline
\end{tabular}
\caption{ \label{tab:1D}
\textbf{Table of Bose-Hubbard parameters for the 1D experiment.}}
\end{table}

\subsection{Higher bands}

As we mention in the main text, we need to keep the drive at small enough frequencies and low enough amplitudes in order to not populate higher bands. Naively, this should only require staying below the gap to the second excited band, $E_\text{g,2}=E_2(q=0)-E_0(q=0) $, since due to symmetry reasons there is no coupling to the first excited band with gap $E_\text{g,1}=E_1(q=\pi/a)-E_0(q=0) $. However, multi-photon resonances can lead to interband transfer even for frequencies well bellow the gap energies, such that in practice we need to identify the regimes at which interband heating starts to take place and stay below those. In Table~\ref{tab:band}, we plot both $E_\text{g,1}$ and $E_\text{g,2}$, also obtained from band-structure numerics, for five different lattice depths within the explored regimes. All the frequencies used to drive the lattice depth in the experiment are well below both $E_\text{g,1}$ and $E_\text{g,2}/3$. 

\begin{table}[h]
\begin{tabular}{|c|c|c|c|c|}
\hline
$V_0$($E_\text{r}$) & $E_\text{g,1}/h$(kHz) &  $E_\text{g,1}  (J)$ &  $E_\text{g,2}/h $(kHz) & $E_\text{g,2} (J)$  \\ \hline
3       & 3.9      & 17 & 9.1   & 40    \\ \hline
5       & 5.5      & 41 & 10.6   & 79    \\ \hline
7       & 7.1      & 89 & 12.5   & 156    \\ \hline
9       & 8.7      & 176 & 14.7  & 298    \\ \hline
11       & 10.1      & 328 & 16.9   & 544    \\ \hline

\end{tabular}
\caption{ \label{tab:band}
\textbf{Table with the bandgaps for different lattice depths.}}
\end{table}

\subsection{Bare residual heating}

In addition to the thermalization processes induced by our well-controlled periodic driving, unwanted bare heating will also take place in the system, even in the absence of any drive. The experiments in this work, however, are performed before a notable onset of these heating mechanisms  (which typically occurs for timescales $> 3000 \, \hbar /J$).

The main decoherence mechanism in our experiment stems from mechanical vibrations and phase noise in the optical-lattice lasers. This mainly leads to intraband heating in the system, giving rise to a slow increase of temperature even when the lattice is not actively modulated. \\

\section{Linear heating in the experiment \label{app:lr_exp}}
\subsection{Heating dynamics}

\begin{figure}
\centering
\includegraphics[width=80mm]{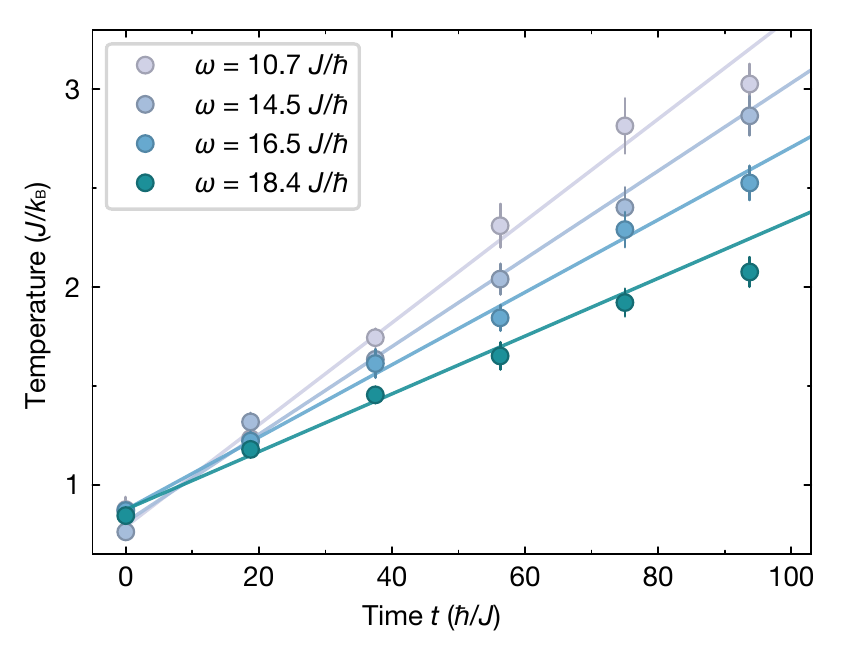}
\caption{ \label{fig:A1}
\textbf{Heating dynamics at $V_0 = 6\, E_\mathrm{r}$.} Temperature of a driven system for four different frequencies as a function of evolution time $t$. The continuous lines are linear fits. The errorbars denote the s.e.m.}
\end{figure}

The experimental heating rates shown in Fig.~\ref{fig:2d} and Fig.~\ref{fig:1d} are extracted from the temperature dynamics within the linear heating regime. The heating rate per Floquet cycle can be expressed as $\phi(\omega)=k_B \, dT/dt \times 2 \pi/\omega$, where $k_B$ is the Boltzmann constant, $T$ the temperature, and $2 \pi/\omega$ is the drive cycle period. We verify that we probe the system within linear response by plotting in Fig.~\ref{fig:A1} sample heating dynamics in 2D at $V_0 = 6 E_\mathrm{r}$ for four different driving frequencies with a relative driving amplitude $A=0.05$. The initial temperature of the cloud is typically around $0.1 \, U_0 / k_B $, where $U_0/h=\SI{660}{Hz}$ is the interaction strength at the atomic limit. In units of the tunneling strength $J$ at ${6 E_\mathrm{r}}$  the temperature is roughly $k_\mathrm{B}T=1.4 J$. We can see that the increase in temperature is consistent with being in the linear response regime.

\subsection{Scaling from Fermi's Golden Rule}

\begin{figure}
\centering
\includegraphics[width=80mm]{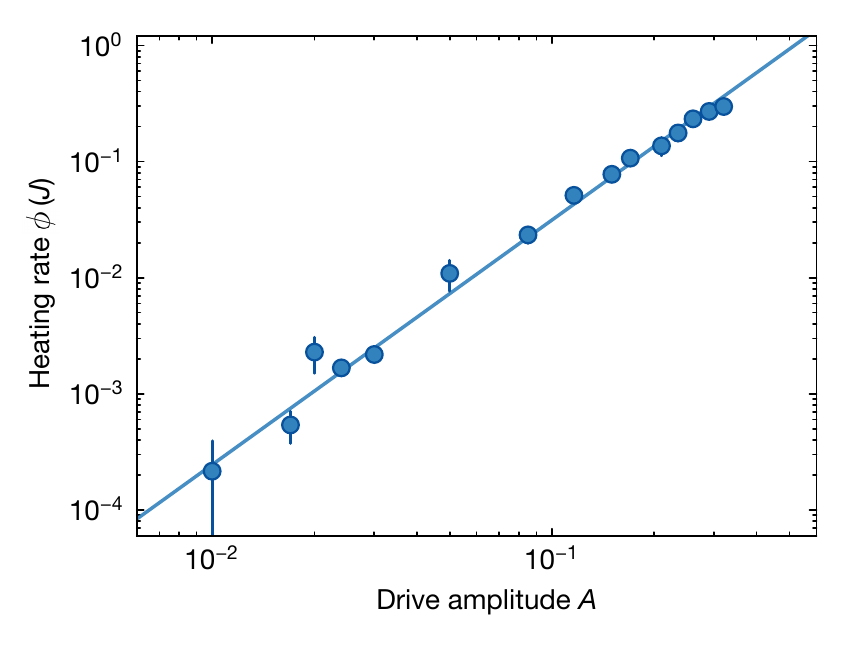}
\caption{ \label{fig:A2}
\textbf{Fermi-golden-rule scaling.} Heating rate as a function of the drive amplitude on a log-log scale. The data was taken at $ V_0=8\, E_\mathrm{r}$ with driving frequency $\omega=14.5 \,J/\hbar$. The solid line is a power-law fit. The errorbars denote the s.e.m. }
\end{figure}

We have also explored the heating rates for different amplitudes of the drive in 2D (see Fig.~\ref{fig:A2}), since in the linear regime we expect a scaling of the heating rate predicted by Fermi's Golden Rule, i.e. proportional to the drive amplitude squared. From fitting a power law with the expression $\phi(A)=cA^\alpha$ we obtain $c=4.0(4)$ and $\alpha=2.11(4)$.

\section{Prethermalization bound for bosons \label{app:ptb}}

The bound on the spectral function (linear-response heating rate) proven in Ref.~\cite{DimaPrethermal_linearresponse}, 
\begin{equation}
\Phi(\omega) < C e^{-\hbar \omega/J_{\rm eff}}
\label{eq:prethermal_bound_supp}
\end{equation}
for appropriate constants $C$ and $J_{\rm eff}$, is strictly speaking only valid for for systems with a bounded local Hilbert space such as spins, fermions or hardcore bosons.
The Bose-Hubbard system we study in this work instead allows for unbounded occupation of each site, which was argued in Ref.~\cite{DimaPrethermal_linearresponse} to generically relax the bound \eqref{eq:prethermal_bound_supp} from exponential to stretched-exponential, 
\begin{equation}
\Phi(\omega) < C e^{-(\hbar \omega/J_{\rm eff})^\alpha}\;, \qquad 0<\alpha<1\;.
\label{eq:relaxed_bound_supp}
\end{equation}
Here we discuss the relevance of this relaxed bound to our experimental observations.

As the physical reason for relaxing the bound to Eq.~\eqref{eq:relaxed_bound_supp} has to do with unbounded energy density, the most natural place to look for violations of the original exponential bound is the Mott-insulating phase.
There, $\Phi(\omega)$ is strongly non-monotonic and exhibits peaks near $\hbar \omega=mU$, $m\in\mathbb N$, as shown in Fig.~\ref{fig:numerics}(b).
These can be understood via perturbation theory from the Mott-insulator state in the atomic limit $J=0$:
each $\hbar \omega = mU$ peak appears at some order $p_m$ in perturbation theory, giving a leading contribution $\sim (J/U)^{2p_m}$, or
$$
\Phi(\omega = mU/\hbar) < C e^{-2p_m \log(U/J)}\;,
$$
in a form reminiscent of the prethermal bounds.

Generally, the optimal process to absorb the most energy in the fewest moves consists of depleting a whole contiguous region on the lattice and gathering all its particles on a central site. 
Taking a sphere of radius $R$ in the hypercubic lattice in $d$ dimensions, this process gives an energy absorption of $\omega \sim R^{2d}$ in $p\sim R^{d+1}$ ``moves'', yielding $p_m\sim (m^{1/2d})^{d+1}$.
This scaling gives in general a stretched exponential bound like Eq.~\eqref{eq:relaxed_bound_supp} with power
$$
\alpha = \frac{d+1}{2d} \;.
$$
In one dimension, we recover $\alpha = 1$, i.e. the exponential bound. 
On the other hand, in higher dimension this construction gives a series of peaks that violate the exponential bound. 
However, such peaks occur at very high frequency and are  sparsely distributed in the spectrum -- in 2D, the lowest such peak is the quintuplon at $\hbar \omega = 10\,U$ (gathering all four nearest neighbors), followed by the ``13-uplon'' at $\hbar \omega = 78\,U$ (gathering the eight next-nearest neighbors), etc. 
Such high frequencies are unlikely to be achievable in experiment without exciting other degrees of freedom, and are thus irrelevant in practice.

In the superfluid phase, single-site high-occupancy excitations play a less prominent role than they do in the Mott phase, which means the above physics is even less likely to be relevant to observations, justifying the use of the exponential bound of Eq.~\eqref{eq:prethermal_bound_supp} somewhat beyond its domain of mathematical rigor.

We conclude by noting that Ref.~\cite{Parker2019} found that the quantity $J_{\rm eff}$ in the exponential bound Eq.~\eqref{eq:prethermal_bound_supp} is fundamentally connected to many important aspects of quantum dynamics, including operator growth and chaos.
This raises interesting questions about such issues in bosonic systems, particularly whether similar bounds on chaos and complexity also hold for a generalized bound like Eq.~\eqref{eq:relaxed_bound_supp}.

\section{Numerical simulations \label{app:numerics}}

\subsection{Details on linear response calculation}

The linear-response expression Eq.~\eqref{eq:LR}, evaluated in a finite-sized system, consists of a finite number of $\delta$ functions.
In order to turn it into a smooth function and plot it, we replace the $\delta$ functions by narrow Gaussians, $\frac{1}{\sqrt{2\pi} \Delta\omega} e^{-\frac{1}{2} (\omega/\Delta\omega)^2}$ (this becomes $\delta(\omega)$ as $\Delta\omega\to0$).
We set $\Delta\omega = 0.01U/\hbar$ and also sample $\omega$ in increments of $\Delta \omega$.

To reduce noise in the resulting data for $\Phi(\omega)$ in Fig.~\ref{fig:numerics}, we additionally perform a moving window average over up to 10 consecutive datapoints.\\

\subsection{Temperature dependence}

The linear-response function in Eq.~\eqref{eq:LR}, and the numerical data in Fig.~\ref{fig:numerics} computed from it, apply to a system in the ground state, i.e. at $T=0$. 
The temperature dependence of this quantity is an interesting subject, especially in some bosonic systems where it has been argued that prethermalization may be a statistical property dependent on the choice of initial state~\cite{DallaTorre2018, DallaTorre2019}.
Here we clarify this issue by computing $\Phi(\omega)$ for different temperatures.

For a finite temperature $k_B T = 1/\beta$, the expression for $\Phi(\omega)$ in Eq.~\eqref{eq:LR} becomes
\begin{align}
\Phi(\omega,\beta) 
& = \sum_{m,n} |\bra{n} \hat{O}_\text{drv} \ket{m} |^2 \,(p_\beta (m)-p_\beta (n)) \nonumber \\
& \qquad \times \delta(E_n-E_m-\hbar\omega)  \;,
\label{eq:LRfiniteT}
\end{align}
where $p_\beta(n)$ is the Boltzmann weight of eigenstate $n$, $p_\beta(n) = e^{-\beta E_n} / \text{tr}(e^{-\beta \hat{H}_0 })$.
In the high-temperature (low $\beta$) limit Eq.~\eqref{eq:LRfiniteT} becomes
\begin{align}
\Phi(\omega,\beta) 
& \simeq \beta \, \frac{\hbar \omega}{\mathcal D} \sum_{m,n} |\bra{n} \hat{O}_\text{drv} \ket{m} |^2 \, \delta(E_n-E_m-\hbar\omega)  \;,
\label{eq:LRhighT}
\end{align}
where $\mathcal D$ is the Hilbert space dimension, $\mathcal D = \binom{2L-1}{L}$ for a chain of $L$ sites at unit filling.
All the temperature dependence in Eq.~\eqref{eq:LRhighT} is captured by the prefactor of $\beta$, which means that the quantity $k_B T \times \Phi$ reaches a temperature-independent value at high $T$. 
This quantity can be computed from the same data used for the $T=0$ case.

\begin{figure}
    \centering
    \includegraphics[width=\columnwidth]{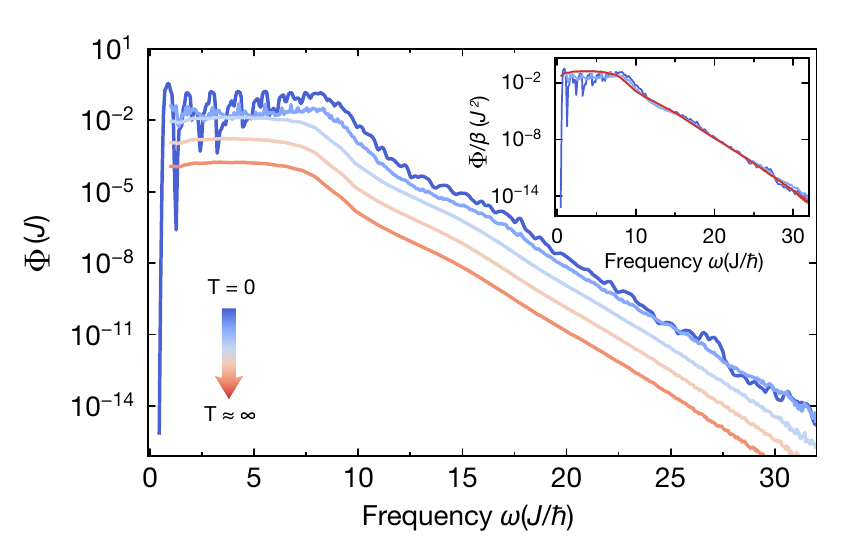}
    \caption{\textbf{Temperature dependence of the linear-response heating rate.} Plot of $\Phi(\omega,\beta) $ as a function of the drive frequency for states at different temperatures. The corresponding temperatures are $k_B T= 0\, J,\,1\, J,\,10\,J,\,100\,J$ and $1000\,J$ (curves from blue to light red). The zero-temperature data is the same shown in Fig.~\ref{fig:numerics} in the superfluid phase ($J/U=2$). In the inset we show the quantity $ k_B T \times \Phi$ for the same temperatures (in the $T=0$ case we plot $\Phi$ directly) and also for the infinite-temperature limit %, given by $\lim_{\,T\to\infty}\, k_B T \times \Phi$ 
    (red curve). The datasets collapse on top of each other, as predicted from Eq.~\eqref{eq:LRhighT}.}
    \label{fig:temperature_dependence}
\end{figure}

In Fig.~\ref{fig:temperature_dependence} we show results obtained for $\Phi(\omega,\beta) $ at finite temperatures together with the corresponding dataset at zero temperature from Fig.~\ref{fig:numerics}. While some of the sharper features visible in the $T=0$ data are smoothed out by the thermal averaging, the main qualitative aspects are preserved.
In particular the exponential suppression remains identical and is in fact even clearer. In the inset of the figure we also take a look at the quantity $k_B T \times \Phi$, which as expected saturates to a finite  %temperature-independent 
value as one moves towards the infinite-temperature limit.

The ``statistical prethermalization'' scenario proposed in Ref.~\cite{DallaTorre2019} for systems of bosons predicts a strong initial-state dependence in heating time and an exponential-in-frequency suppression in the heating rate tuned by an effective temperature, not a fixed local energy scale of the problem.
These behaviors are not seen in the present case. 
One reason for this may be the fact that we consider systems at unit filling, as opposed to the high-occupancy semiclassical limit studied in Ref.~\cite{DallaTorre2019}.

\subsection{Dynamics simulations}

Here we present numerical simulations that complement those of the main text and corroborate their validity at finite drive amplitude.
We use a Krylov subspace method to simulate the dynamics of a state $|\psi(N_\text{cyc})\rangle$ at stroboscopic times $t=N_\text{cyc}\,T$ ($T=2\pi/\omega$ is the drive period) and track the value of the ``energy'' $\langle \psi(N_\text{cyc}) | \hat{H}_0 | \psi(N_\text{cyc})\rangle \equiv E_{N_\text{cyc}}$ during the evolution to define a thermalization time $N_\text{cyc}^\text{th}$. 
This method can probe slightly larger sizes than full diagonalization (up to $L=13$ sites, Hilbert space dimension in the millions) and, most importantly, is not restricted to drive amplitudes $g\ll 1$, at the expense of a limited dynamic range in $N_\text{cyc}^\text{th}$.

\begin{figure}
    \centering
    \includegraphics[width=\columnwidth]{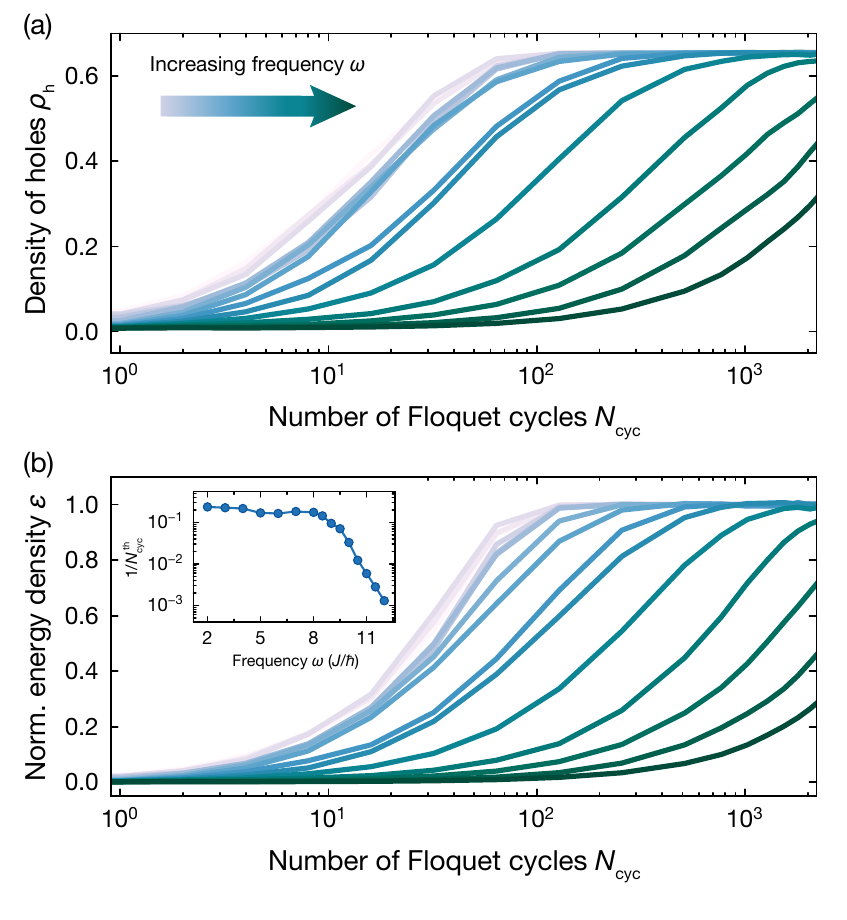}
    \caption{\textbf{Time evolution simulations.}
    One-dimensional chain of $L=12$ sites at unit filling in the superfluid phase, driven with amplitude $g=0.5$ and subsequently ramped to the atomic limit. 
    (a) Density of holes $\rho_\text{h}$ and (b) normalized energy density $\varepsilon$ as a function of the number of Floquet cycles $N_\text{cyc}$ for different values of the frequency of the drive. The frequencies range from $\omega=2\,J/\hbar$ (gray) to $\omega=12\,J/\hbar$ (dark green). Inset: heating rate (inverse number of Floquet cycles $N_\text{cyc}$ where $\varepsilon$ crosses the threshold value $\varepsilon^\star = 0.1$) vs drive frequency $\omega$.}
    \label{fig:dynamics}
\end{figure}

We define a normalized energy density 
$$
\varepsilon(N_\text{cyc}) \equiv \frac{E_{N_\text{cyc}}-E_0}{E_\infty - E_0}\;,
$$
where $E_\infty \propto \text{Tr}(\hat{H}_0)$ is the infinite-temperature value of the energy.
During the dynamics, this obeys $0\leq \varepsilon(N_\text{cyc}) \leq 1$.
We define the heating time $N_\text{cyc}^\text{th}$ as the lowest $N_\text{cyc}$ such that $\varepsilon(N_\text{cyc})$ is greater than some predefined threshold $\varepsilon^\star$ (we use $0.1$, though other choices give similar results).
We also keep track of the ``density of holes''\;,
$$
\rho_\text{h}(N_\text{cyc}) = \frac{1}{L} \sum_{\textbf{\textit{i}}} \langle \psi(N_\text{cyc}) | \hat{\rho}_{\text{h}, \textbf{\textit{i}}} | \psi(N_\text{cyc})\rangle \;,
$$
where $\hat{\rho}_{\text{h}, \textbf{\textit{i}}}$ is a projector onto even occupation of site $\textbf{\textit{i}}$, which mimics the fluorescence imaging technique used in the experiment.
We choose the initial state $|\psi(0)\rangle$ as the ground state of $\hat{H}_0$ (obtained via the Lanczos method). 
We then approximate each Floquet cycle by a sequence of $s$ constant Hamiltonians, $\{\hat{H}(t=Tk/s):\ k=0,\dots s-1\}$, and time-evolve the state vector for time $T/s$ with each of these Hamiltonians using the Krylov subspace method. 
In practice, we use $s=32$ steps; further increasing $s$ changes the results negligibly.\\

\begin{figure}
    \centering
    \includegraphics[width=80mm]{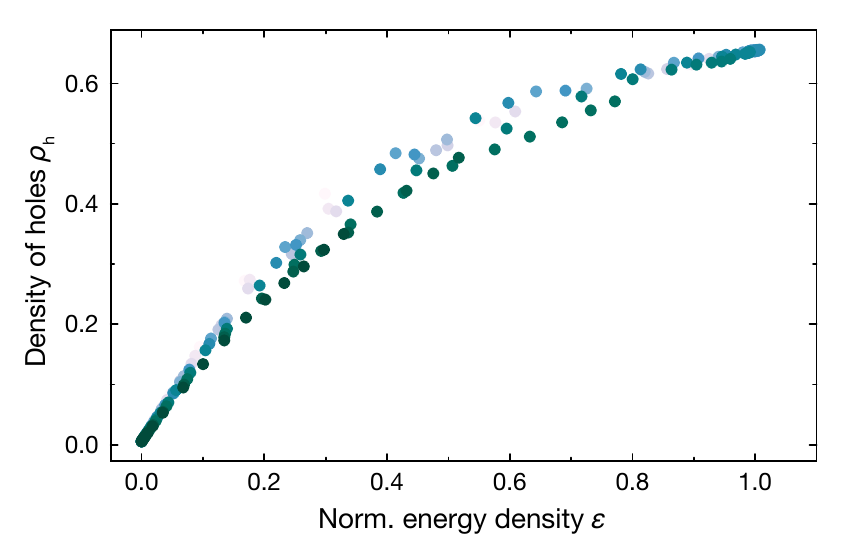}
    \caption{\textbf{Simulated density of holes vs energy density.}
    One-dimensional chain of $L=12$ sites at unit filling in the superfluid phase, driven with amplitude $g=0.5$ and subsequently ramped to the atomic limit. 
    Density of holes $\rho_\text{h}$ plotted as a function of the normalized energy density $\varepsilon$ for different values of the frequency of the drive. The frequencies range from $\omega=2\,J/\hbar$ (gray) to $\omega=12\,J/\hbar$ (dark green).}
    \label{fig:HolevsEnergy}
\end{figure}

In addition, to better imitate the experimental procedure, before measuring the observables $\varepsilon$ and $\rho_\text{h}$ we ramp the system into an atomic-limit Mott insulating state, i.e. we arrest the drive and slowly take $J\to 0$ as $J(N_\text{cyc}T + t) = (1-t/\tau)\,J(N_\text{cyc}T)$ for $0\leq t\leq \tau$, where $\tau$ is a long timescale (we use $\tau = 100\,\hbar/U$). 
In practice, this is again accomplished by time-evolving with piecewise constant Hamiltonians, keeping the same time step used during the drive.
A copy of the wavefunction at $t=N_\text{cyc}\, T$ (before the ramp) is stored so that the time evolution can resume after the measurement is taken.

Results for a fairly strong drive amplitude $g=0.5$ (see Fig.~\ref{fig:dynamics}) generally agree with the linear response picture -- $N_\text{cyc}^\text{th}(\omega)$  is approximately flat for $\omega \lesssim \Omega_{2\text{qp}}$, then starts increasing exponentially.
The time traces of the density of holes $\rho_\text{h}$, shown in Fig.~\ref{fig:dynamics}(a), are very similar to the experimental data in Fig.~\ref{fig:1}(c).
Comparison with the energy density traces in Fig.~\ref{fig:dynamics}(b) also confirms that $\rho_\text{h}$ is indeed a good proxy for the energy density $\varepsilon$.
We further confirm the relation between these two quantities by plotting the density of holes vs the normalized energy density (see Fig.~\ref{fig:HolevsEnergy}).

\begin{figure}
\centering
\includegraphics[width=\columnwidth]{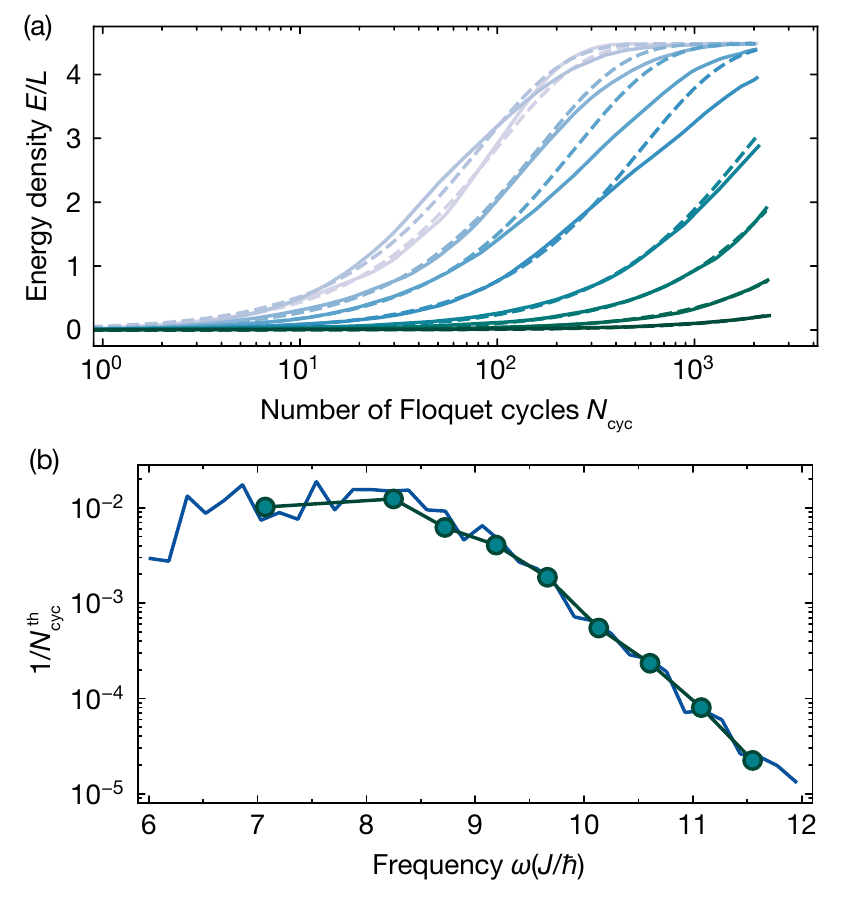}
\caption{\textbf{Comparison between time evolution simulations and linear response theory.}
(a) Energy density vs time from Krylov time evolution simulation of a chain of length $L=12$ (solid lines) and fits to exponentials $E(N_{\rm cyc}) = E_\infty (1-e^{-N_{\rm cyc}/N_{\rm cyc}^{\rm th}})$ (dashed lines). The drive amplitude is $g=0.5$. The frequencies of the rive range from $\omega=7\, J/\hbar $ (gray) to $\omega=11.5 \,J/\hbar $ (dark green).
(b) Inverse time constants $1/N_{\rm cyc}^{\rm th}$ obtained from the fits (green markers) compared with the linear response rate $(\pi g)^2 \Phi(\omega)/E_\infty$ (blue line).
\label{fig:comparison}}
\end{figure}

Finally, we can compare the results of the Krylov time evolution simulations to those of the linear response function in the main text, evaluated via exact diagonalization.
In Fig.~\ref{fig:comparison}(a) we show time evolution data analogous to Fig.~\ref{fig:dynamics}, and fit each curve to a single-timescale exponential, $E(N_{\rm cyc})  = E_\infty (1- e^{-N_{\rm cyc}/N_{\rm cyc}^{\rm th}})$, in an early-time window $E(N_{\rm cyc}) < \frac{1}{4} E_\infty$. 
The best-fit values of $1/N_{\rm cyc}^{\rm th}$ are then shown in Fig.~\ref{fig:comparison}(b) superimposed with the linear-response rate $(\pi g)^2 \Phi(\omega)/E_\infty$ computed via exact diagonalization. 
(Both quantities are intensive and thus we can compare them across different system sizes, $L=12$ for Krylov time evolution and $L=9$ for exact diagonalization.)
The agreement is excellent, even at a quantitative level, and corroborates the validity of the linear-response prediction even at a fairly large value of the coupling ($g=0.5$ in this case).
We remark that the late-time dynamics close to infinite temperature can deviate quite significantly from the early-time prediction, as expected; nonetheless, the timescale $N_{\rm cyc}^{\rm th}$ is a good estimate of the overall heating time -- i.e., by the time the deviation is significant, the prethermal `plateau' has already faded away.

As the method simulates time evolution directly, its cost scales as $O(N_\text{cyc}^\text{th})$, hence exponentially in $\omega$;
this limits us to $\hbar \omega \simeq 12J$, and in particular prevents us from testing the presence of an additional threshold feature near $2\,\Omega_{2\text{qp}}$ as seen in linear response (Fig.~\ref{fig:numerics}(a)).
It also makes the method generally less suited to the Mott-insulating phase, where $N_\text{cyc}^\text{th}$ has non-monotonic oscillations by many orders of magnitude.

\bibliography{FloquetPrethermal_biblio}

\end{document}